\documentclass[aps,prb,preprint,superscriptaddress]{revtex4-1}

\usepackage{times}
\usepackage[T1]{fontenc}
\usepackage[latin1]{inputenc}
\usepackage{epsfig}
\usepackage{graphicx}
\usepackage{graphics}
\usepackage{bm}
\usepackage{hyperref}
\usepackage{amssymb}
\usepackage{amsmath}
\usepackage[english]{babel}
\usepackage{subfig}
\usepackage{color} 

\newcommand{\vvs}{\mathbf{v}_s}
\newcommand{\vvn}{\mathbf{v}_n}
\newcommand{\rr}{\mathbf{r}}

\newcommand{\der}[2]{\frac{\partial {#1}}{\partial {#2}}}

\newcommand{\be}{\begin{equation}}
\newcommand{\ee}{\end{equation}}
\newcommand{\dsp}{\displaystyle}
\newcommand{\beqn}{\begin{eqnarray}}
\newcommand{\eeqn}{\end{eqnarray}}
\newcommand{\refeq}[1]{(\ref{#1})}

\begin{document}


\title{Coupled normal fluid and superfluid profiles of turbulent
helium II in channels}


\author{Luca Galantucci}
\email[]{luca.galantucci@newcastle.ac.uk}
\affiliation{Joint Quantum Centre (JQC) Durham--Newcastle, and
School of Mathematics and Statistics, Newcastle University,
Newcastle upon Tyne, NE1 7RU, United Kingdom}

\author{Michele Sciacca}
\email[]{michele.sciacca@unipa.it}
\affiliation{Dipartimento di Scienze Agrarie e Forestali, Universit\`a di Palermo}

\author{Carlo F. Barenghi}
\email[]{carlo.barenghi@newcastle.ac.uk}
\affiliation{Joint Quantum Centre (JQC) Durham--Newcastle, and
School of Mathematics and Statistics, Newcastle University,
Newcastle upon Tyne, NE1 7RU, United Kingdom}


\date{\today}

\begin{abstract}
We perform fully coupled two--dimensional numerical simulations
of plane channel helium II counterflows with vortex--line density typical
of experiments. The main features of our approach are the inclusion of 
the back reaction of the superfluid vortices on the normal fluid and the
presence of solid boundaries. Despite the reduced dimensionality, our model
is realistic enough to reproduce vortex density distributions across the channel
recently calculated in three--dimensions. 
We focus on the coarse--grained superfluid and normal fluid velocity profiles, recovering the 
normal fluid profile recently observed employing a technique based on laser--induced 
fluorescence of metastable helium molecules.
\end{abstract}

\pacs{\{67.25.dk\}, \{47.37.+q\}, \{47.27.nd\}}

\maketitle

\section{Introduction\label{sec: Intro}}

Three--dimensional homogeneous isotropic turbulence is the
benchmark of turbulence research.  Recent papers 
\cite{skrbek-sreenivasan-2012,nemirovskii-2013,
barenghi-skrbek-sreenivasan-2014} have compared the properties of
homogeneous isotropic turbulence in
ordinary (classical) fluids and in liquid helium near 
absolute zero, and found remarkable similarities.  
In particular, experiments have revealed that
the temporal decay of vorticity \cite{stalp-skrbek-donnelly-1999} 
is the same, and that the energy spectrum 
(which represents the distribution of kinetic energy over the length scales)
obeys the same classical Kolmogorov scaling at sufficiently 
large length scales,
\cite{maurer-tabeling-1998,salort-etal-2010,
barenghi-lvov-roche-2014}
in agreement with theoretical 
\cite{lvov-nazarenko-skrbek-2006,lvov-nazarenko-rudenko-2008} 
and numerical studies.
\cite{baggaley-laurie-barenghi-2012,
baggaley-barenghi-2011,araki-tsubota-nemirovskii-2002,
kobayashi-tsubota-2005} These results are surprising, because
the low temperature phase of $^4$He (hereafter referred to simply
as helium~II), is quite different from an ordinary fluid.
\cite{donnelly-1991} It is well--known, in fact, that
helium~II consists of two interpenetrating fluid 
components: a viscous normal fluid  (whose
vorticity is unconstrained) and an inviscid superfluid 
(whose vorticity is confined to vortex line singularities of fixed 
circulation $h/m$ where $h$ is Planck's constant and 
$m$ the mass of one helium atom).  

Despite its importance, isotropic homogeneous turbulence is an idealization
which neglects the role of boundaries (for example, vorticity is generated
at the walls of a channel).  In this report we are concerned with
superfluid turbulence along channels or pipes.
Such flows are neither homogeneous (because the boundary conditions are
likely to induce non--uniform profiles) nor isotropic (because of
the direction of the flow). The prototype
channel problem of the helium literature is thermal counterflow.
\cite{vinen-1957a,vinen-1957c,brewer-edwards-1961,childers-tough-1976,ladner-tough-1979,
martin-tough-1983,tough-1982} 
The typical experimental set--up consists of
a channel which is closed at one end, and is open to the helium bath at the
other end. At the closed end, a resistor dissipates a known
heat flux which is carried away by the normal fluid; to conserve
mass, superfluid flows in the opposite direction towards the resistor; 
the resulting velocity difference between the two fluids is
proportional to the applied heat flux.  If this heat flux is
larger than a small critical value,
the superfluid component becomes turbulent, forming 
a disordered tangle of quantised vortex lines (superfluid turbulence). 
The intensity of
the vortex tangle is usually characterized by
its vortex line density $L$ (length of quantized
vortex lines per unit volume), which can be determined by
measuring the attenuation of second sound as a function of the
applied heat flux.

The questions which we address in this work
is simple but fundamental: what are the profiles
of the normal fluid, of the superfluid, and of the vortex density 
across the counterflow channel?

This question motivates current experimental attempts to directly
visualize the flow of helium~II. Two new visualization methods stand out.
Particle Tracking Velocimetry (PTV) of hydrogen and/or deuterium flakes
\cite{bewley-lathrop-sreenivasan-2006,chagovets-vansciver-2011,lamantia-chagovets-rotter-skrbek-2012} 
has been used to image individual quantum vortex reconnections 
\cite{bewley-paoletti-sreenivasan-lathrop-2008}
and to determine the velocity and acceleration statistics of the 
turbulent superfluid. \cite{paoletti-etal-2008,lamantia-duda-rotter-skrbek-2013b}
Laser--induced fluorescence of metastable helium molecules 
\cite{guo-etal-2010,marakov-etal-2015} 
has directly imaged the profile of the normal component,
addressing the issue of whether, at sufficiently large heat currents,
the normal fluid flow undergoes a laminar--turbulent transition. 
\cite{melotte-barenghi-1998}

Until now, the question of the profiles of normal fluid, superfluid
and vortex line density has been unanswered.
On first thoughts, in analogy with a classical viscous fluid (which obeys 
the Navier--Stokes equation with no--slip boundary conditions), the normal 
fluid component should have a parabolic Poiseuille profile across the
channel; similarly,
in analogy with a classical inviscid fluid
(which obeys the Euler equation and, unimpeded by viscosity, can slip along
the channel's walls), the superfluid
component should have a uniform profile, hence the vortex line density
should be non polarized and, eventually, uniform. 
On second thoughts, the said profiles cannot be
correct: the varying mismatch
between the superfluid and normal fluid velocities across the channel
would induce a large non--uniform mutual friction 
\cite{barenghi-donnelly-vinen-1983}
which would modify these profiles. To appreciate the
mathematical difficulty of the problem, notice that not only are the two
fluid components coupled, (the normal fluid affects
the superfluid and viceversa), but the coupling term between the
two fluids is nonlinear:
the mutual friction is proportional to the velocity difference between
normal fluid and superfluid, times the vortex line density, which is
a nonlinear function of this velocity difference. 

Unfortunately, most numerical simulations of superfluid turbulence 
in the literature have determined the superfluid vortex tangle
in the presence of a {\it prescribed} normal fluid, without taking into
account the back reaction of the vortex lines on the normal fluid.
Various models of the imposed normal fluid have been studied: uniform,
\cite{schwarz-1988,adachi-fujiyama-tsubota-2010,baggaley-sherwin-barenghi-sergeev-2012,sherwin-barenghi-baggaley-2015}  
parabolic,\cite{aarts-dewaele-1994,baggaley-laizet-2013,khomenko-etal-2015,baggaley-laurie-2015} 
Hagen--Poiseuille and tail--flattened flows,\cite{yui-tsubota-2015} 
vortex tubes,\cite{samuels-1993} ABC flows,\cite{barenghi-samuels-bauer-donnelly-1997} 
frozen normal fluid vortex tangles \cite{kivotides-2006}, 
random waves \cite{sherwin-barenghi-baggaley-2015},
time--frozen snapshots of the turbulent solution of 
the Navier--Stokes equations
\cite{baggaley-laizet-2013,sherwin-barenghi-baggaley-2015,baggaley-laurie-2015} and 
time--dependent homogeneous and isotropic turbulent solutions of linearly forced
Navier--Stokes equations.\cite{morris-koplik-rouson-2008} 
Moreover,
most calculations were performed
in open or periodic domains, avoiding the difficulty of the boundary.
Other works have determined the
effects of a prescribed superfluid tangle on the normal fluid,
\cite{melotte-barenghi-1998} failing again to fully model the coupling
of superfluid and normal fluid. Because of the computational complexity
and cost involved, fully coupled calculations have been attempted only for 
simple configurations, such as single, isolated vortex lines
\cite{idowu-willis-barenghi-samuels-2000}
or rings,
\cite{kivotides-barenghi-samuels-2000} 
or for decaying tangles in open geometry \cite{kivotides-2011}
and periodic domains.\cite{kivotides-2007,kivotides-2014} 

The model which we present here is fully coupled (the normal fluid affects
the superfluid and viceversa via a nonlinear mutual friction term) 
and includes boundaries.
To cope with the computational difficulty, our model
is two--dimensional rather than three-dimensional: vortex
loops in a three-dimensional channel are thus replaced by vortex points in a 
two--dimensional channel. 
Despite the simplified dimensionality,
our model captures the nonlinearity of the problem,
which, we think, is the key ingredient to determine flow profiles
in actual channels.

The outline of the paper is the following.
In Section~\ref{sec: Method} we describe the 
two--dimensional model which we use and the details of the numerical algorithm.
Section~\ref{sec: num sim} focuses on the results and in
Section~\ref{sec: Discussion} we critically discuss to what extent our two--dimensional
model is capable of grasping the most relevant vortex dynamics occurring in
helium II counterflows. 
Finally, Section~\ref{sec: Conclusions} summarizes the conclusions.

\section{Model\label{sec: Method}}

\subsection{The counterflow channel\label{subsec: count.channel}}

We consider an infinite two--dimensional channel of width $D$. 
Let $x$ and $y$ be respectively the directions along and across the channel
with walls at $y=\pm D/2$ and periodic boundary conditions imposed at $x=0$ 
and $x=L_x$.  The average normal fluid and superfluid flows are respectively
in the negative and positive $x$ direction. 

The superfluid vortices are modelled as $N$ vortex--points of 
circulation $\Gamma_j$ and position $\rr_j(t)=\left ( x_j(t) , y_j(t) \right )$, where $j=1,\cdots N$ and
$t$ is time.

Half the vortices have positive circulation $\Gamma_j=\kappa$ and half have negative circulation $\Gamma_j=-\kappa$, 
where $\kappa=10^{-3} \rm cm^2/s$ is the quantum of circulation in 
superfluid $^4$He. 

To make connection with experiments we interpret $n=N/(DL_x)$ 
(average number of vortex--points per unit area) as the two--dimensional 
analogue of the 
three--dimensional vortex--line density $L$,
and relate $L$ to the channel--averaged normal fluid longitudinal velocity 
$\langle u_n \rangle$ via the relation \cite{ladner-tough-1979} 

\begin{equation}
L^{1/2}D=1.03 \gamma_0 \frac{\rho}{\rho_s} \langle u_{n} \rangle h_D - 1.48\beta,
\label{eq:L-Vns}
\end{equation}

\noindent
where $\langle u_n\rangle$ is related to the applied heat flux $q$ via

\begin{equation}
\langle u_n \rangle = \frac{q}{T \rho S},
\end{equation}

\noindent
where $T$ is the absolute temperature, $S$ the specific entropy, 
and $\rho_n$, $\rho_s$ and $\rho$ the normal fluid, superfluid and total
helium II densities, respectively, where $\rho=\rho_n+ \rho_s$. The 
coefficients  $\gamma_0$ and $\beta$ in Eq. \refeq{eq:L-Vns}
have been determined 
experimentally by Tough and collaborators,
\cite{ladner-tough-1979,tough-1982,martin-tough-1983} 
while $h_D$ represents the channel's hydraulic diameter.  

In the absence of vortices, the counterflow condition of zero net mass flow
\begin{equation}
\dsp
\rho_n \langle u_{n} \rangle + \rho_s v_s^{ext} = 0 \, , \label{eq:count.cond.vortex.free}
\end{equation}
determines the uniform superflow $v_s^{ext}$ in the opposite direction with respect to the 
normal fluid.
Notice that Eq. \refeq{eq:L-Vns} coincides with Vinen's equation \cite{vinen-1957c}
describing the evolution of the vortex--line density $L$ modified in order
to take into account the presence of solid boundaries and that the 
average intervortex distance $\ell$ is defined by the relation $\ell=L^{-1/2}$.

\subsection{The superfluid vortices\label{subsec: vortices}}

The vortex points move according to
\cite{schwarz-1988}

\begin{eqnarray}
\displaystyle
\frac{d\rr_j}{dt} & = & \vvs (\rr_j,t) + \alpha \, \mathbf{s}_j' \times \left (\vvn(\rr_j,t) - \vvs(\rr_j,t) \right ) \nonumber \\[2mm] 
&& + \alpha' \left (\vvn(\rr_j,t) - \vvs(\rr_j,t) \right )\label{eq:r_j}
\end{eqnarray} 

\noindent
where $\mathbf{s}_j'$ is the unit vector along vortex $j$ (in the positive 
or negative $z$ direction depending on whether $\Gamma_j$ is positive
or negative), $\alpha$ and $\alpha'$ are temperature
dependent mutual friction coefficients \cite{barenghi-donnelly-vinen-1983},
$\vvn(\rr_j,t)$ is the normal fluid velocity at position $\rr_j$;
the superfluid velocity at position $\rr_j$ is decomposed as
\begin{equation}
\vvs(\rr_j,t)= \mathbf{v}_s^{ext}(t)+\mathbf{v}_{si}(\rr_j,t), \label{eq: v_s=vsi+vsext}
\end{equation}

\noindent
where
$\mathbf{v}_s^{ext}(t)$ 
is the uniform (potential) superfluid  
flow which enforces the counterflow condition of no net mass flow and
$\mathbf{v}_{si}(\rr_j,t)$ is the superfluid velocity field induced by 
all the $N$ vortex--points at $\rr_j$:

\begin{equation}
\dsp
\mathbf{v}_{si}(\rr_j,t) = \sum_{k=1 \dots N} \mathbf{v}_{si,k}(\rr_j,t) \, .
\end{equation}

The integration in time of Eq. \refeq{eq:r_j} is performed employing the
second--order Adams--Bashfort temporal advancement scheme.

To determine the superfluid velocity field induced by the $k$-th vortex 
$\mathbf{v}_{si,k}(\mathbf{x},t)$ we employ a complex--potential--based 
formulation enforcing 
the boundary condition that, at each wall, the superfluid has zero
velocity component in the wall--normal direction.

The complex potential can be derived using conformal 
mapping \cite{saffman-1992} or, equivalently, using (for each vortex) an
infinite number of images with respect to the channel walls,
\cite{greengard-1990} leading to the following expression

\beqn\label{eq:F_j}
\dsp
F_k(z,t) = \mp i \frac{h}{2\pi m}\log \frac{\sinh \left [ \frac{\pi}{2D} (z - z_k(t)) \right ]}{\sinh \left [ \frac{\pi}{2D} (z - \overline{z_k}(t) ) \right ] }
\eeqn

\noindent
where $z_k(t)=x_k(t) +iy_k(t)$ is the complex number associated 
to $\rr_k(t)$. The corresponding superfluid velocity
$\mathbf{v}_{si,k}(z,t)=\left ( v_{si,k}^x,v_{si,k}^y \right )$
is obtained from the complex potential in the usual way as 

\begin{equation}
v_{si,k}^x  - iv_{si,k}^y = \frac{dF_k(z,t)}{dz}
\end{equation}

The uniform superfluid velocity
$\mathbf{v}_s^{ext}(t)=\left ( v_s^{ext}(t), 0 \right )$ in 
Eq. \refeq{eq: v_s=vsi+vsext} is instead obtained by enforcing at each timestep the 
counterflow condition of no net mass flow taking into account the presence of vortices, \textit{i.e}

\beqn\label{eq:counterflow_condition}
\dsp
\rho_n \langle u_{n} \rangle + \rho_s \left ( \langle u_{si} \rangle (t) + v_s^{ext}(t) \right ) = 0 \,\, .
\eeqn

\noindent
where $\mathbf{v}_{si}=\left ( u_{si} , v_{si} \right )$ to ease notation. 

To model the creation and the destruction of vortices (mechanisms 
intrinsically three-dimensional) within our two--dimensional model, 
we proceed as follows. When the
distance between two vortex points of opposite circulation becomes 
smaller than a
critical value $\epsilon_1$,  we perform a ''numerical vortex reconnection'' 
and remove these
vortex points; similarly, when the distance between a vortex point and a 
boundary is
less than $\epsilon_2 = 0.5\epsilon_1$, we remove this vortex point 
(the vortex of opposite circulation
being the nearest image vortex beyond the wall). 
To maintain a steady state, when a
vortex point is removed, a new vortex point of the same circulation 
is re-inserted into
the channel in a random position.  
In order to assess the dependence of the numerical results on the value of 
$\epsilon_1$, we have performed numerical simulations varying the value of 
$\epsilon_1$ by two orders of magnitude: we find that the results are identical. 
This reconnection model, corresponding 
three--dimensionally to the
vortex filament method of Schwarz,\cite{schwarz-1988}  
correctly describes the fate of two very near
antiparallel vortices (as confirmed by past Gross-Pitaevskii numerical 
studies \cite{koplik-levine-1993})
and avoids the generation of infinitesimal length scales which would 
trigger numerical instabilities.
In order to estimate the impact of this re--insertion procedure on the numerical results, 
another two--dimensional renucleation model has also been explored in the 
present study: the vortices are re--inserted
with the same wall--normal coordinate $y$ with which they have been
removed and a random streamwise $x$ coordinate. The results obtained are quasi--identical to the ones obtained with the
random re--insertion model, concluding that the numerical results 
presented in Section \ref{subsec: results}, referring to the random re--insertion model, 
are not an artificial outcome of the reconnection procedure.


\subsection{The normal fluid\label{subsec: vn}}

Typical experimental values of pressure and temperature variations 
along counterflow 
channels allow us to assume that both superfluid and normal fluid flows are
incompressible and isoentropic, \textit{i.e.} 
$\rho\, , \rho_n \, , \rho_s \, , S$ are constant.  
Furthermore, assuming negligible the variations of the normal fluid dynamic 
viscosity $\eta_n$ and of the thermal conductivity $\lambda$ across the channel 
and neglecting quadratic or higher--order terms
in spatial gradients of velocity and thermodynamics variables, the 
resulting incompressible and isoentropic equations of motion of
the normal fluid are the following:\cite{landau-1941,bekarevich-khalatnikov-1961} 

\beqn\dsp
 \der{\vvn}{t} + \left ( \vvn\cdot\nabla \right )\vvn & = &-\frac{1}{\rho}\nabla p -\frac{\rho_s}{\rho_n} S \nabla T + \nu_n \nabla^2 \vvn \nonumber \\
&&\nonumber \\ 
&& - \frac{\rho_s}{2\rho}\nabla \left ( \vvn - \vvs \right )^2 + \frac{1}{\rho_n}\widetilde{\mathbf{F}}_{ns}\label{eq: vn counterflow}
\eeqn
\be\dsp
 \nabla\cdot\vvn  =  0 \label{eq: div vn =0}
\ee
where $\nu_n=\eta_n/\rho_n$ is the normal fluid kinematic viscosity, 
and the mutual friction force
$\widetilde{\mathbf{F}}_{ns}$ is determined by the averaging procedure described
in Section~\ref{subsec: Fns}.

The normal fluid velocity field $\vvn$ is decomposed 
in two solenoidal fields:
\begin{equation}
\vvn = \vvn^p +\vvn'.
\end{equation}

\noindent
The first field 
$\vvn^p=\left ( u_n^p, v_n^p \right ) = \left ( -V_{n0} \left [ 1- (2y/D)^2 \right ], 0 \right )$ 
is the Poiseuille flow which would exist in absence of superfluid vorticity 
at constant heat flux $q$ supplied by the heater. 
The second velocity field $\vvn'=\left ( u_n', v_n' \right )$ accounts 
for the back reaction of the superfluid vortex--lines
on the normal fluid. To calculate $\vvn'$ we employ the vorticity--stream 
function formulation, according to which we define the stream function
$\Psi'$ and vorticity field $\omega_n'$ as follows:
\beqn\dsp
\vvn'&=&\left ( \der{\Psi'}{y}, -\der{\Psi'}{x} \right )\label{eq: v_n'-psi}\, , \\[5mm]
\omega_n'&=&\left ( \nabla\times \vvn'\right )\cdot \hat{\mathbf{z}}\, ,\label{eq: w_n'-v_n'}
\eeqn
where $\hat{\mathbf{z}}$ is the unit vector in the $z$ direction.
The definition of $\Psi'$, Eq. \refeq{eq: v_n'-psi}, directly ensures that $\vvn'$ is solenoidal, Eq. \refeq{eq: div vn =0}, 
while the Navier--Stokes equations \refeq{eq: vn counterflow} are equivalent to the following two scalar equations:
\beqn\dsp
\nabla^2\Psi' = -\omega_n'\label{eq: poisson psi}
\eeqn
\beqn\dsp 
\der{\omega_n'}{t}+\left ( u_n^p + \der{\Psi'}{y} \right )\der{\omega_n'}{x} &-&\der{\Psi'}{x}\left ( \der{\omega_n'}{y} - \frac{d^2u_n^p}{dy^2}  \right ) = \nonumber \\[5mm]
\nu_n \nabla^2 \omega_n' + &\dsp \frac{1}{\rho_n}& \left ( \der{\widetilde{F}^y}{x} - \der{\widetilde{F}^x}{y} \right )  \label{eq: omega_n.3}
\eeqn
where $\widetilde{\mathbf{F}}_{ns}=\left ( \widetilde{F}^x , \widetilde{F}^y \right )$. 

The evolution equation \refeq{eq: omega_n.3} for the normal vorticity $\omega_n'$ is discretized in space
employing second--order finite differences and its temporal integration 
is accomplished using the second--order Adams--Bashfort numerical scheme.
The Poisson equation \refeq{eq: poisson psi} is instead solved in a mixed $(k_x,y)$ space, 
employing a Fourier--spectral discretization in the periodic 
$x$--direction and second--order finite differences in the wall--normal direction $y$. 
The boundary conditions on $\Psi'$ and $\omega_n'$ are deduced 
by imposing no--slip boundary conditions on the viscous normal fluid velocity field.

\subsection{The mutual friction\label{subsec: Fns}}

The mutual friction force $\mathbf{F}_{ns}$ accounts for the momentum exchange between the normal fluid and the superfluid 
in presence of the quantized vortex--lines which act as scattering centres for the elementary excitations constituting
the normal component.\cite{hall-vinen-1956b} This exchange takes place at very small length--scales, less than the average intervortex
distance $\ell$, beyond the practical numerical resolution and, at some temperatures, the hydrodynamical description of the normal
fluid. To make progress, we employ the coarse--grained theoretical framework elaborated by Hall and Vinen \cite{hall-vinen-1956b}
according to which, at lengthscales larger than $\ell$, the mutual friction forcing assumes 
the following expression  
\be
\dsp\widetilde{\mathbf{F}}_{ns}=\alpha\rho_s\widehat{\widetilde{\bm{\omega}}}_s \times \left [ \widetilde{\bm{\omega}}_s \times \left ( \widetilde{\mathbf{v}}_n - \widetilde{\mathbf{v}}_s \right ) \right ] + \alpha' \rho_s\widetilde{\bm{\omega}}_s \times \left ( \widetilde{\mathbf{v}}_n - \widetilde{\mathbf{v}}_s \right )\,\, , \label{eq:F_ns}
\ee  
where $\widetilde{\, \cdot \,}$ symbols indicate coarse--grained averaged quantities. 

We distinguish between the $\left ( \Delta x, \Delta y \right )$ grid on which the normal fluid velocity $\vvn$ is numerically determined, and
the coarser $\left (\Delta X,\Delta Y \right )$ grid on which we define the mutual friction $\widetilde{\mathbf{F}}_{ns}$. In principle, we would like 
to have $\Delta X$ and $\Delta Y~\gg~\ell $ corresponding to the Hall--Vinen limit; in practice, we use $\Delta X$ and $\Delta Y~>~\ell $ due to computational constraints.
To prevent rapid fluctuations of the friction at small length--scales, we smooth the vortex distribution using the Gaussian kernel $\Theta_j(\mathbf{r})$
associated to each vortex $j$ according to the following expression 

\beqn\dsp
\Theta_j(\mathbf{r})= \frac{1}{V_j} \dsp e^{\dsp-\frac{|\mathbf{r}-\mathbf{r}_j|^2}{2\ell^2}} \, , 
\eeqn
where $\dsp V_j=\!\!\!\int\limits_{0}^{L_x}\!\int\limits_{-D/2}^{D/2}\!\!\!e^{-\frac{|\mathbf{r}-\mathbf{r}_j|^2}{2\ell^2}} dxdy$. 
Hence, on the basis of Eq. \refeq{eq:F_ns}, the mutual friction force $\widetilde{\mathbf{F}}_{ns}^{p,q}$ averaged on the coarse grid--cell $(p,q)$ is given by the following expression
\beqn\dsp
\widetilde{\mathbf{F}}_{ns}^{p,q} & = & -\alpha \rho_s\, \kappa L^{p,q} \left ( \widetilde{\mathbf{v}}_n^{p,q} - \widetilde{\mathbf{v}}_s^{p,q} \right ) \nonumber \\[5mm]
&& + \alpha' \rho_s \,\Omega^{p,q}\,\hat{\mathbf{z}}\times\left (\widetilde{\mathbf{v}}_n^{p,q} - \widetilde{\mathbf{v}}_s^{p,q}\right )\label{eq:Fns^pq}
\eeqn
where
\beqn\dsp
L^{p,q} &=&\sum_{j=1 \dots N}\,\, \frac{1}{\Delta X \Delta Y} \iint\limits_{(p,q)} \Theta_j(\mathbf{r}) d\mathbf{r} \label{eq: L^pq}\\[5mm]
\Omega^{p,q} &=& \sum_{j=1 \dots N}\,\, \frac{\Gamma_j}{\Delta X \Delta Y} \iint\limits_{(p,q)} \Theta_j(\mathbf{r}) d\mathbf{r} \label{eq: omega^pq}
\eeqn
$\Gamma_j=\pm\kappa$ and the symbol $\dsp\iint\limits_{(p,q)}$ 
denotes the integral over the coarse grid--cell $(p,q)$. 
Physically, $\dsp L^{p,q}$ corresponds to the coarse--grained 
vortex--line density while $\dsp \Omega^{p,q}$ coincides 
with the coarse--grained superfluid vorticity.
Finally, we average $\widetilde{\mathbf{F}}_{ns}^{p,q}$ over the short
time interval $T_{ns}=\Delta X/v_s^{ext}$, the average time 
interval during which a vortex--point moves from a coarse grid--cell
to the neighbouring (cfr. Eq. \refeq{eq:r_j}). 

\begin{figure}[htbp]
 \begin{center} 
\hspace{-6cm}    
  \includegraphics[width=0.5\textwidth]{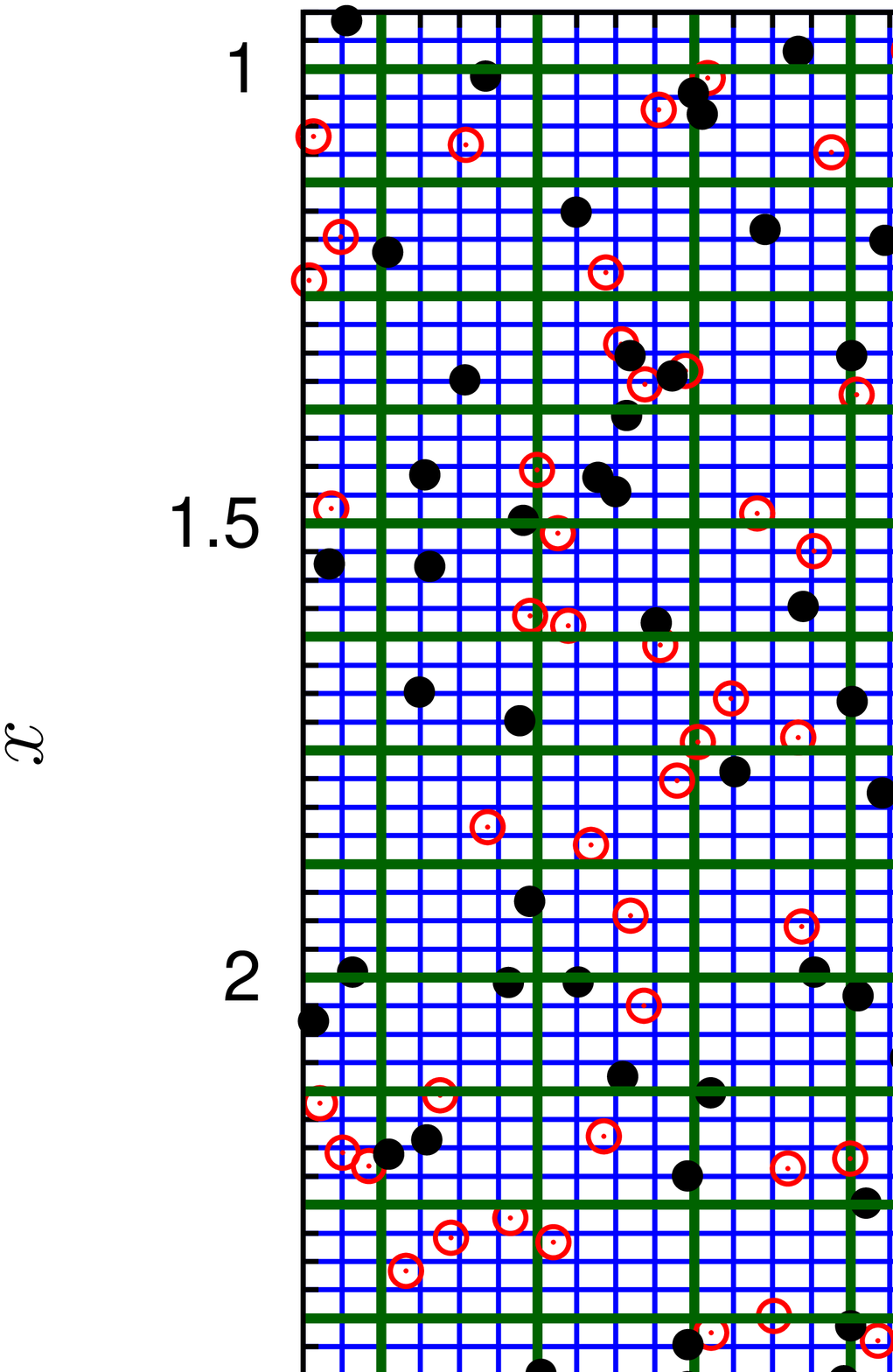}   
\vspace{2.0cm}
  \caption{(Color online). The structure of the fine (blue solid lines) 
and coarse grids (green solid lines) are illustrated on a particular portion
of the computational domain, together with positive and negative
vortices indicated with empty red and filled black circles, respectively. 
  \label{fig: coarse-graining_1}}         
 \end{center}
\end{figure}

\begin{figure}[htbp]
 \begin{center}
\vspace{-6cm} 
\hspace{-5.cm}
  \includegraphics[width=0.55\textwidth,height=1.09\textwidth]{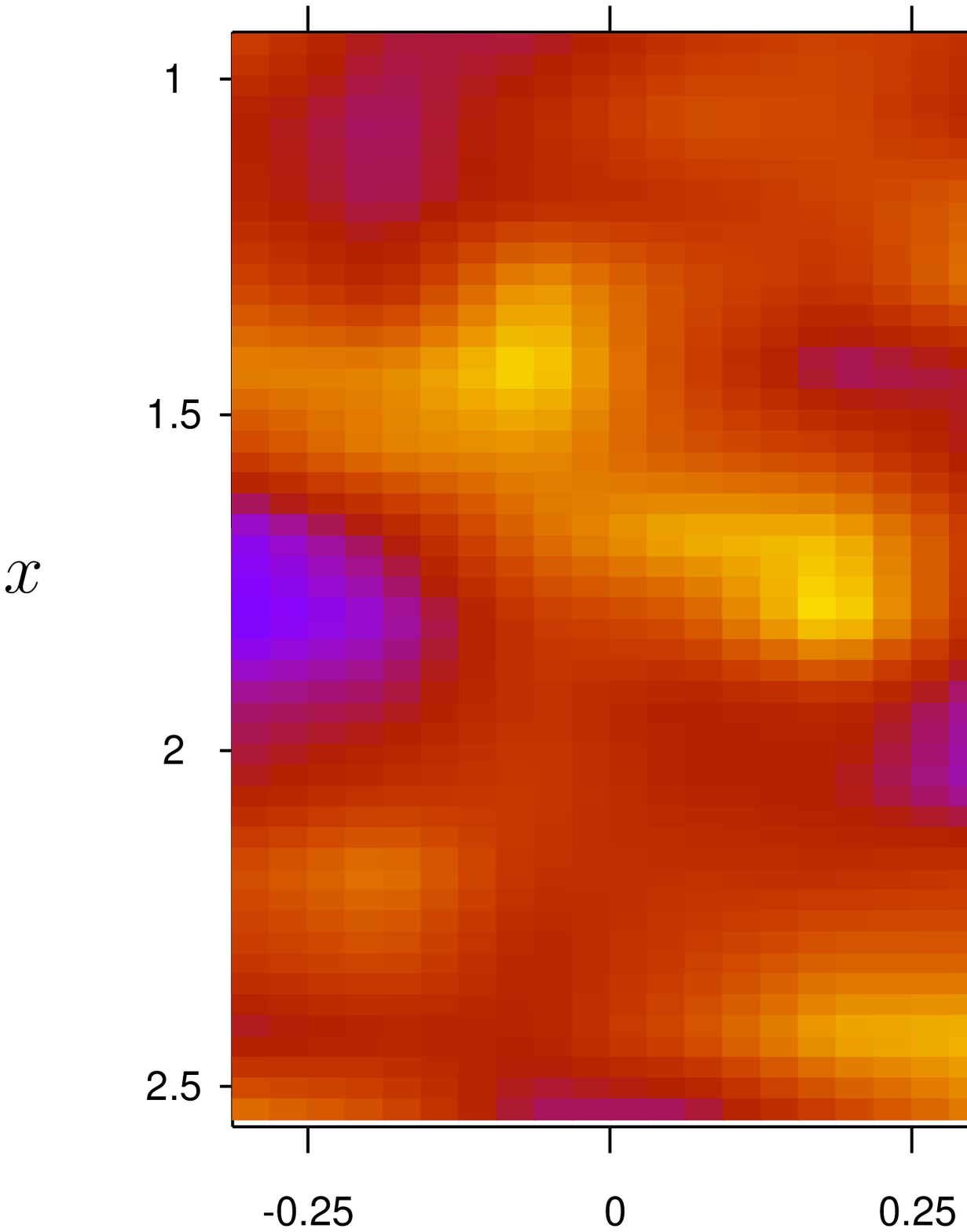}     
\vspace{2.0cm}
  \caption{(Color online). Two--dimensional color plot of the longitudinal component of the 
mutual friction force $\widetilde{F}^x/\rho_n$ (see Eq. \refeq{eq: omega_n.3}),
interpolated on the fine grid on the
same domain as Fig. \ref{fig: coarse-graining_1}. 
The axes of the plot are rescaled employing the scaling units 
defined in  Section \ref{subsec: param}.
  \label{fig: coarse-graining_2}}
 \end{center}
\end{figure}

The interpolation of $\widetilde{\mathbf{F}}_{ns}$ on the 
finer grid $\left ( \Delta x, \Delta y \right )$ is performed
via a two--dimensional bi--cubic convolution kernel \cite{keys-1981} 
whose order of accuracy is between linear interpolation and cubic splines orders of accuracy.
The structure of the fine and coarse grids on a particular 
portion of the computational domain is illustrated in 
Fig. \ref{fig: coarse-graining_1}, while 
in Fig. \ref{fig: coarse-graining_2} we report a two--dimensional 
color plot of the longitudinal component of the 
mutual friction force $\widetilde{F}^x$ interpolated on the fine grid, on the
same domain as Fig. \ref{fig: coarse-graining_1}: 
the smoothing effect of the Gaussian kernel combined with the interpolating 
scheme emerges clearly, if compared to the ideally 
$\delta$--shaped nature of $\mathbf{F}_{ns}$ centered on the vortex--points
displayed in Fig. \ref{fig: coarse-graining_1}.
Furthermore, it is worth emphasizing that the employment of 
Eq. \refeq{eq:Fns^pq} for the computation of the mutual 
friction force $\widetilde{\mathbf{F}}_{ns}$, 
ensures a smooth transition when the vortex--points 
cross coarse grid--cell boundaries. 

\section{Numerical simulations\label{sec: num sim}}

\subsection{Parameters\label{subsec: param}}
We chose the parameters of the numerical simulations in order to be able to make at least qualitative comparisons with 
experiments. As a reference, we select the experimental counterflow studies performed by Tough and collaborators on both
high aspect--ratio rectangular cross--section channels, \cite{ladner-tough-1979} which represent the closest real experimental 
settings to our idealized plane channel, and cylindrical capillary tubes \cite{childers-tough-1976,martin-tough-1983}. 
More in detail, we set the width of the channel $D=9.1\times 10^{-3}\,\rm cm$, corresponding to tube R4 in 
Ref.~[\onlinecite{ladner-tough-1979}], and $n^{1/2}D=25$. The consequent Reynolds number of the normal fluid flow 
calculated via Eq. \refeq{eq:L-Vns} is $Re=206$, far below the critical Reynolds
number for the onset of classical turbulent channel flows $Re_c \approx 5772$.
\cite{orszag-1971} As a consequence, on the basis also of past experimental investigations,
\cite{childers-tough-1976,ladner-tough-1979,martin-tough-1983} we reckon that in our
numerical experiment the flow of the normal fluid is still laminar.

\begin{table}
\begin{tabular}{|c|c||c|c|}
\hline
$\,\,\,\,\, D \,\,\,\,\,$ & $2$ & $V_{n0}$ & $553.6$ \\ 
\hline
$L_x$ & $6$ & $T$ & $1.7 K$ \\
\hline
$N$ & $1876$ & $\rho_s/\rho_n$ & $3.373$ \\
\hline
$n$ & $\,\,\,\,\,\,\, 156.3 \,\,\,\,\,\,\,$ & $\epsilon_1$ & $2.5\times 10^{-3}$ \\
\hline
$\ell$ & $0.08$ & $\Delta t_v$ & $\,\,7.5\times 10^{-6} \,\, $\\
\hline
\end{tabular}
\caption{Numerical parameters employed in the simulations\\ and subsequent physical relevant quantities
in dimensionless units}
\label{table: param}
\end{table}

The complete list of parameters employed in our simulation and the subsequent physical relevant quantities 
are reported in Table \ref{table: param}, expressed in terms of the following units of length, velocity and time, respectively:
$\delta_c=D/2=4.55\times 10^{-3}\,\rm cm$, $u_c=\kappa/(2\pi\delta_c)=3.49\times 10^{-2}\,\rm cm/s$, 
$t_c=\delta_c/u_c=0.13\, \rm s$. Hereafter all the quantities which we mention 
are dimensionless, unless otherwise stated. The constant $V_{n0}$ determining 
$u_n^p$ is computed imposing, without any loss of generality, 
that the whole normal fluid flow rate is supplied by the Poiseuille field $\vvn^p$, \textit{i.e.}
$\langle u_n \rangle=\langle u_n^p \rangle$, implying $\langle u_n' \rangle=0$. In the spirit of 
the coarse--grained description illustrated in Section \ref{subsec: Fns}, we define a coarse 
and a fine grid characterized by numbers of grid--points and spacings listed in Table \ref{table: grid}, satisfying the condition 
$\Delta X, \Delta Y > \ell > \Delta x, \Delta y$. 

\begin{table}
\begin{tabular}{||c|c||c|c||}
\hline
\multicolumn{2}{||c||}{Fine grid} & \multicolumn{2}{|c||}{Coarse grid} \\
\hline \hline
$\,\, n_x \,\,$ & $192$ & $\,\, N_x \,\, $ & $48$ \\ 
\hline
$n_y$ & $64$ & $N_y$ & $16$ \\
\hline
$\Delta x$ & $\,\,3.125\times 10^{-2}\,\,$ & $\Delta X$ & $0.125$ \\
\hline
$\Delta y$ & $\,\,3.125\times 10^{-2}\,\,$ & $\Delta Y$ & $\,\, 0.125 \,\, $ \\
\hline
\end{tabular}
\caption{Number of grid--points and spacings in dimensionless units \\
of the grids employed in the numerical simulations}
\label{table: grid}
\end{table}	

The coupled calculation of vortex motions and $\vvn$
entails the simultaneous existence of two different timestep stability criteria, one for each motion. Concerning the 
evolution equation \refeq{eq: omega_n.3} for $\omega_n'$, the constraint is set by the normal fluid viscosity,\cite{peyret-taylor-1983} 
leading to the restriction $\Delta t_n\le (\Delta x)^2/\nu$. Regarding the motion of the superfluid vortices,
consistently with the numerical reconnection procedure illustrated in Section \ref{subsec: vortices}, the integration timestep $\Delta t_v$
for Eq. \refeq{eq:r_j} must satisfy the condition $\Delta t_v\le \epsilon_1/(2V_{\epsilon_1})$, where $V_{\epsilon_1}$ is the velocity 
of a pair of anti--vortices along their separation vector when separated by a distance equal to $\epsilon_1$. This constraint on $\Delta t_v$ 
prevents from the generation of unphysical small--scale periodic motions (\textit{e.g.} vortex--pairs multiple crossings). 
The value of $\Delta t_v$ employed in our simulation is reported in Table \ref{table: param} and 
the viscous constraint allows us to set $\Delta t_n=2\Delta t_v$, implying
that vortex motions alternately take place with frozen normal fluid.
 

\subsection{Results\label{subsec: results}}

\subsubsection{Steady--state regime\label{subsubsec: eq}}

\begin{figure}[htbp]
\vspace{-1.0cm}
     \hspace{-3.0cm}
     \begin{minipage}{\textwidth}
      \centering
       \includegraphics[width=0.285\textwidth,height=0.35\textwidth]{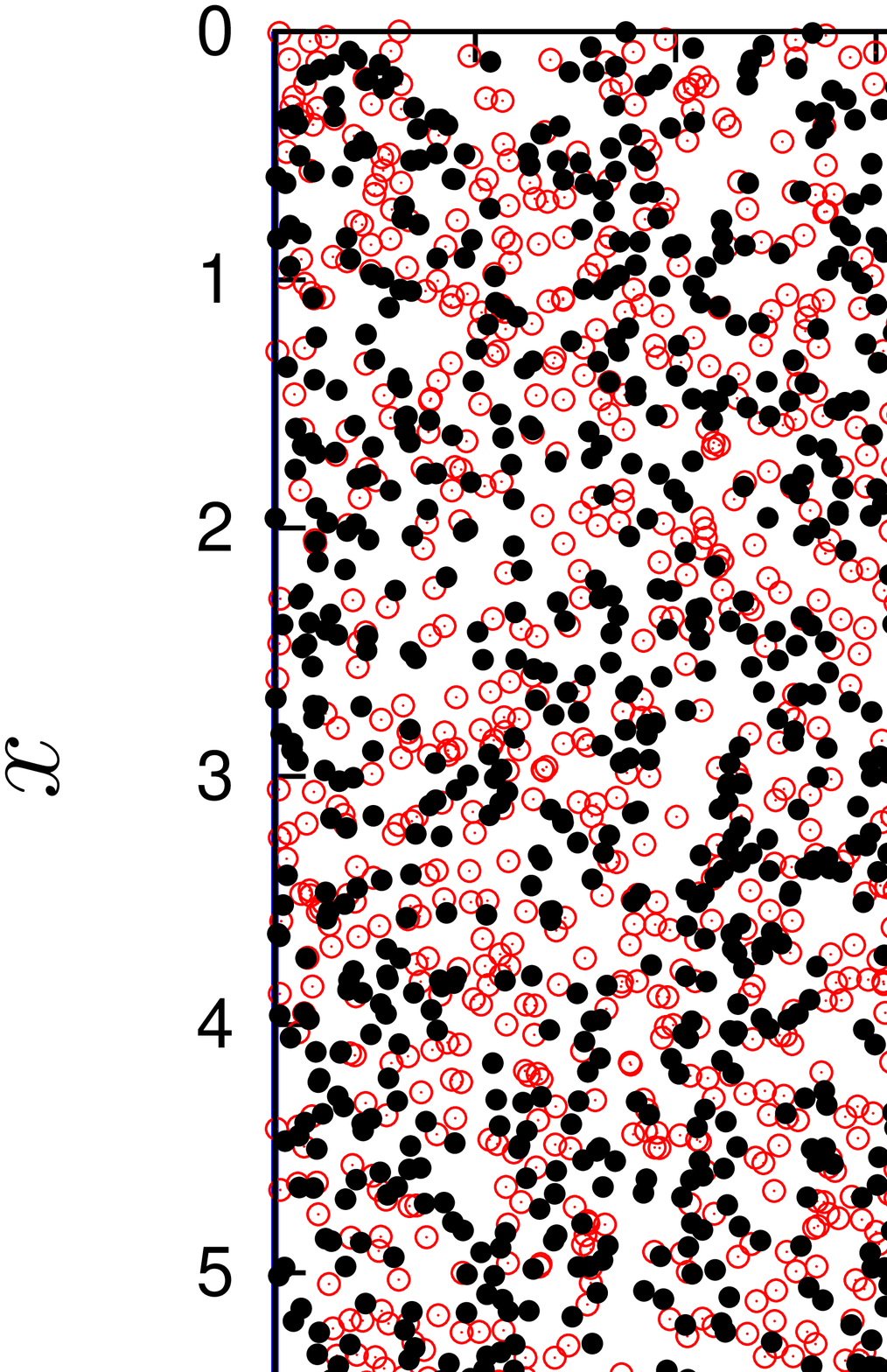}     
     \end{minipage}\\[5mm]
    \hspace{-3.0cm}
     \begin{minipage}{\textwidth}
      \centering
       \includegraphics[width=0.285\textwidth,height=0.35\textwidth]{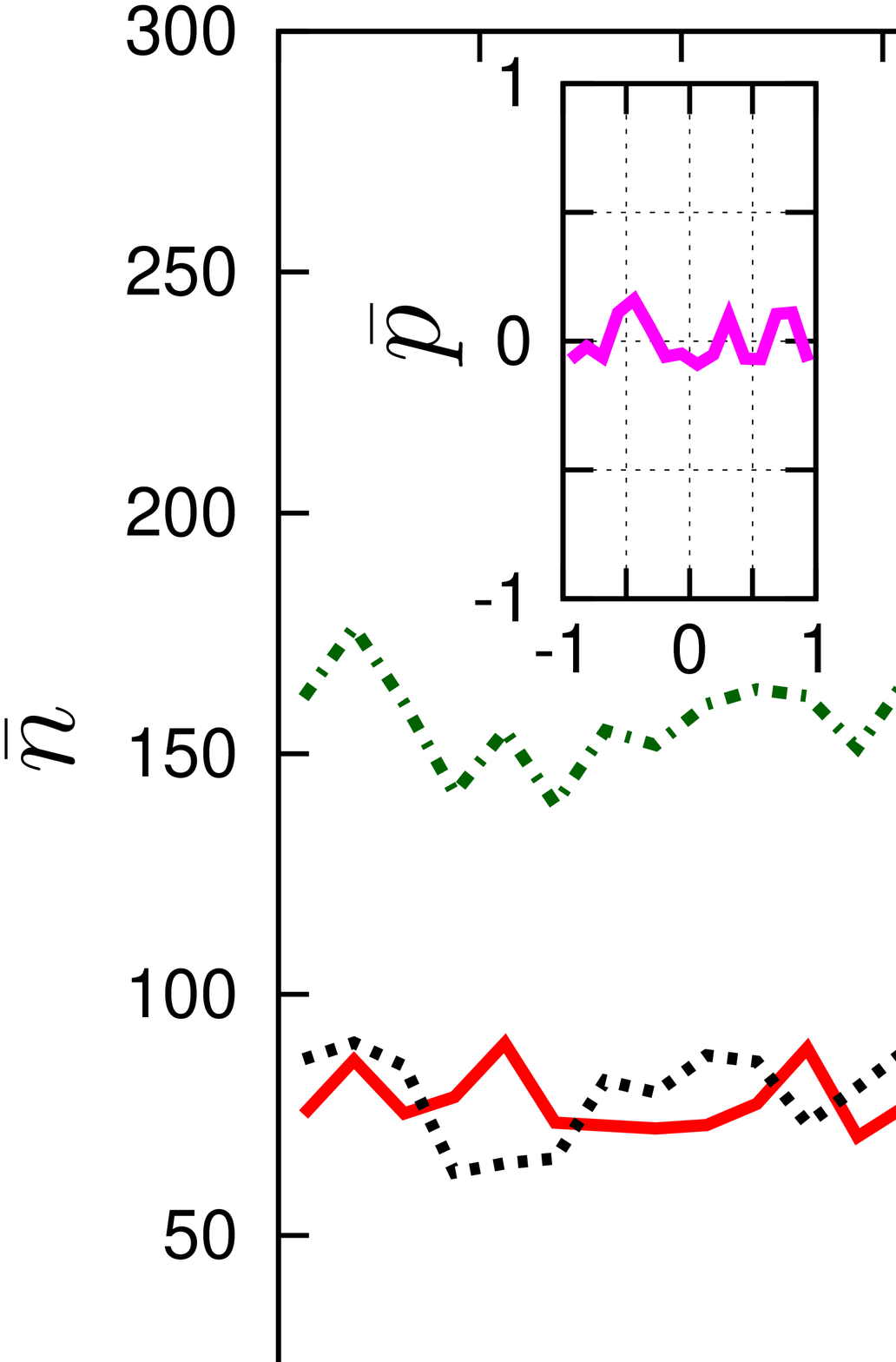}       
     \end{minipage}\\[5mm]
     \hspace{-3.0cm}
     \begin{minipage}{\textwidth}
      \centering
       \includegraphics[width=0.285\textwidth,height=0.35\textwidth]{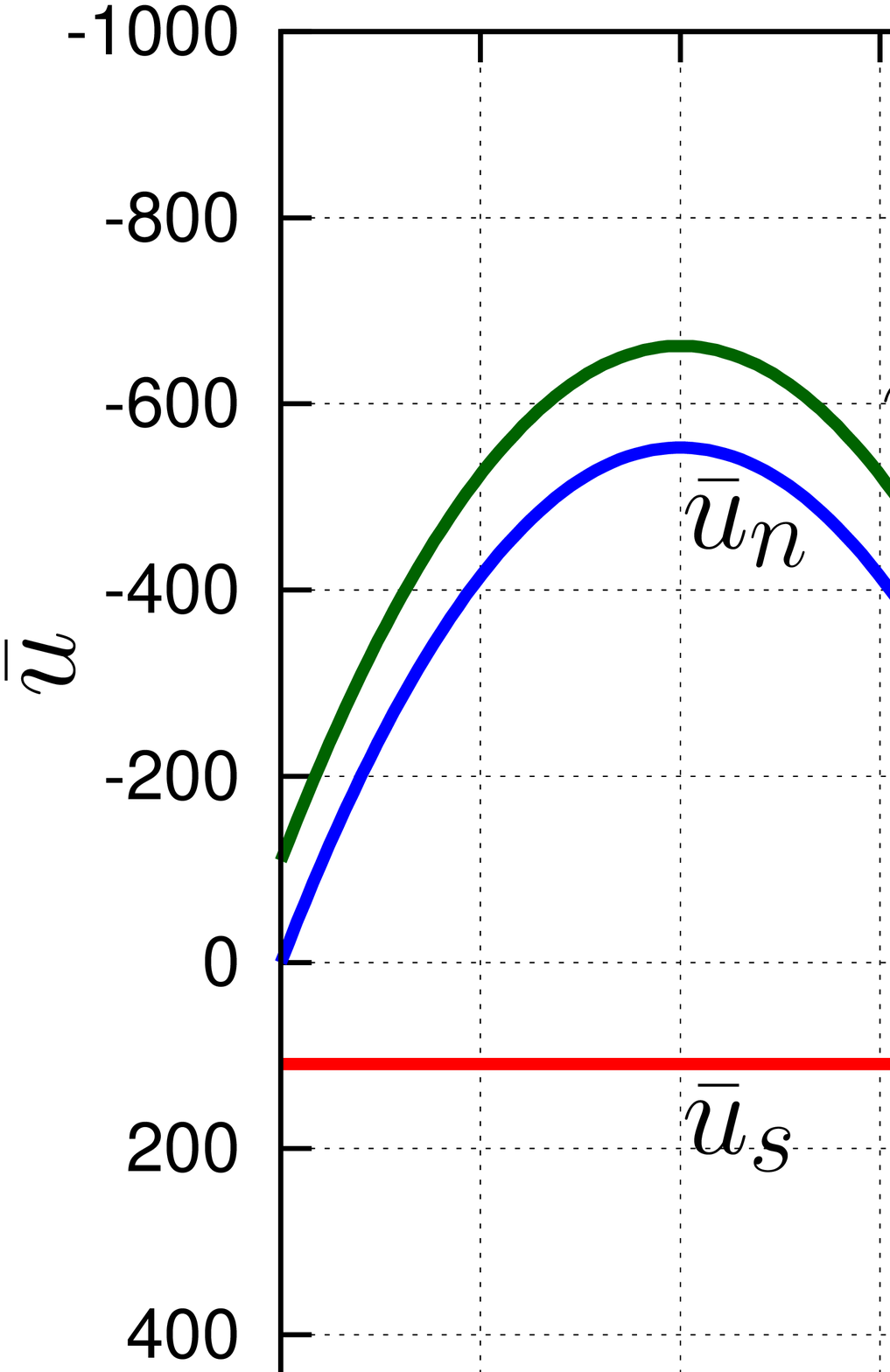}            
     \end{minipage}
\vspace{1.0cm}
\caption{(Color online). (top): vortex distribution at $t=0$, red empty (black filled) 
circels indicate positive (negative) vortices; 
(middle): coarse--grained profiles of positive vortex density 
$\overline{n}^+$ (solid red line), negative vortex density 
$\overline{n}^-$ 
(dashed black line) and total vortex density $\overline{n}$ 
(dot--dashed green line) at $t=0$. In the inset, the corresponding 
coarse--grained profile of the polarization magnitude $\overline{p}(y)$ is reported (solid magenta line); 
(bottom) coarse--grained profiles of superfluid velocity
$\overline{u}_s$ (solid red line), normal fluid velocity $\overline{u}_n$ 
(solid blue line) and counterflow velocity $\overline{u}_{ns}=\overline{u}_{n}-\overline{u}_{s}$ (solid green line)
at $t=0$ \label{fig:t=0}}
\end{figure}

The aim of our numerical simulations is to determine the spatial 
distributions of positive and negative vortices and the normal
fluid and superfluid velocity profiles across the channel in the 
steady--state regime. To stress that these distributions and profiles
are meant to be coarse--grained over channel stripes of size $\Delta Y$,
we use the $\overline{\,\cdot\,}$ symbols. Fig.~\ref{fig:t=0} 
illustrates the initial conditions of a typical simulation. 
Fig.~\ref{fig:t=0} (top) shows the initial random 
spatial distribution of the vortices, corresponding 
to the coarse--grained vortex density profiles 
$\overline{n}(y)$ shown in Fig.~\ref{fig:t=0} (middle). 
In Fig.~\ref{fig:t=0} (bottom) the initial parabolic Poiseuille 
profile for $\dsp\overline{u}_n$ and the flat 
profile for $\dsp\overline{u}_s$ are reported.
After a transient interval whose characteristics will be addressed 
in section \ref{subsubsec: trans}, the system reaches the statistically--steady--state 
described in Fig.~\ref{fig:t=Teq}. As expected, the steady--state 
regime is achieved after a time interval $T_f\approx D^2/\nu$. 
The most important feature is the shape of the 
coarse--grained profile of the normal fluid velocity $\dsp\overline{u}_n$ reported in Fig.~\ref{fig:t=Teq} (bottom), 
which is slightly flattened in the near--wall region and sharpened 
in the central region with respect to the Poiseuille profile. 
These characteristics have recently been observed experimentally
by means of laser--induced fluorescence \cite{marakov-etal-2015} 
in the same counterflow regime (turbulent superfluid,
laminar normal fluid). In the experiment, the flattening of the profile is more 
pronounced, but we reckon that this difference is due, at least partially, 
to a larger superfluid turbulent 
intensity in the experimental setting ($40 \lesssim L^{1/2}D \lesssim 70$ against 
$L^{1/2}D\simeq 25$ in our simulation).
 
The other key feature which emerges from the numerical simulation is 
the polarization of the superfluid vortex distribution, which can be 
qualitatively observed in 
the snapshot of the steady--state vortex configuration, 
Fig.~\ref{fig:t=Teq} (top).
To investigate quantitatively this aspect, we introduce
the coarse--grained polarization vector $\overline{\mathbf p}(y)$ 
defined by \cite{Jou-Sciacca-Mongiovi-2008}
\be 
\displaystyle 
\overline{\mathbf{p}}(y)=\frac{\overline{\bm{\omega}}_s(y)}{\kappa \overline{n}(y)}=\frac{\overline{n}^+(y)-\overline{n}^-(y)}{\overline{n}^+(y)+\overline{n}^-(y)}\hat{\mathbf{z}}\, .
\ee 
Note that $\overline{\mathbf p}(y)=\mathbf{0}$
when quantum turbulence is uniformly distributed  
all over the channel (as, for instance, at $t=0$ in our numerical simulations,
see Fig.~\ref{fig:t=0} (top) and (middle) ).
The steady--state profile of the polarization magnitude $\overline{p}(y)$
is reported in Fig.~\ref{fig:t=Teq} (middle) together with  the positive and negative 
vortex density profiles, $\overline{n}^+(y)$ and $\overline{n}^-(y)$ respectively.
This polarized pattern directly arises from 
the vortex--points equations of motion \refeq{eq:r_j}, where the friction term 
containing $\alpha$ depends on the polarity of vortex. 

This polarization of the vortex configuration, which, we stress, is 
\emph{not} complete, \emph{i.e.} $|\overline{p}(y)|<1$,
generates a parabolic coarse--grained 
superfluid velocity profile $\overline{u}_s(y)\sim y^2$ 
which is reported in Fig.~\ref{fig:t=Teq} (bottom).
This process, \emph{i.e.} the superfluid polarization induced by a normal fluid shear 
generating a superfluid velocity pattern which mimics the normal fluid one,
confirms past analytical results obtained via simple models \cite{barenghi-hulton-samuels-2002}
and backs numerically observed 
normal fluid--superfluid velocity matching and vorticity locking.
\cite{samuels-1993,barenghi-samuels-bauer-donnelly-1997,barenghi-hulton-samuels-2002,morris-koplik-rouson-2008} 

It is interesting to notice that our model, although being two--dimensional, recovers the 
total vortex density profile $\overline{n}(y)$ 
computed very recently via three--dimensional numerical simulations of helium II channel counterflows 
with prescribed Poiseuille normal flow.\cite{baggaley-laizet-2013,khomenko-etal-2015,baggaley-laurie-2015} 
On the contrary, the
vortex density profile $\overline{n}(y)$ calculated in this work is significantly different from the 
ones computed in past two--dimensional simulations with prescribed Poiseuille normal flow,
where the density is approximately uniform across the channel.
\cite{galantucci-barenghi-sciacca-etal-2011,galantucci-sciacca-2012} 

\subsubsection{Transient interval\label{subsubsec: trans}}

The main results of our investigations have been outlined in the previous section. 
Before we finish, it is instructive to describe how the vortices and the normal 
fluid adjust to each other reaching a steady--state, starting from our arbitrary
initial condition: this exercise helps to understand the physics of the coupling
of vortices and normal fluid. 

\begin{figure}[htbp]
\vspace{-2.0cm}
     \hspace{-3.0cm}
     \begin{minipage}{\textwidth}
      \centering
       \includegraphics[width=0.285\textwidth,height=0.35\textwidth]{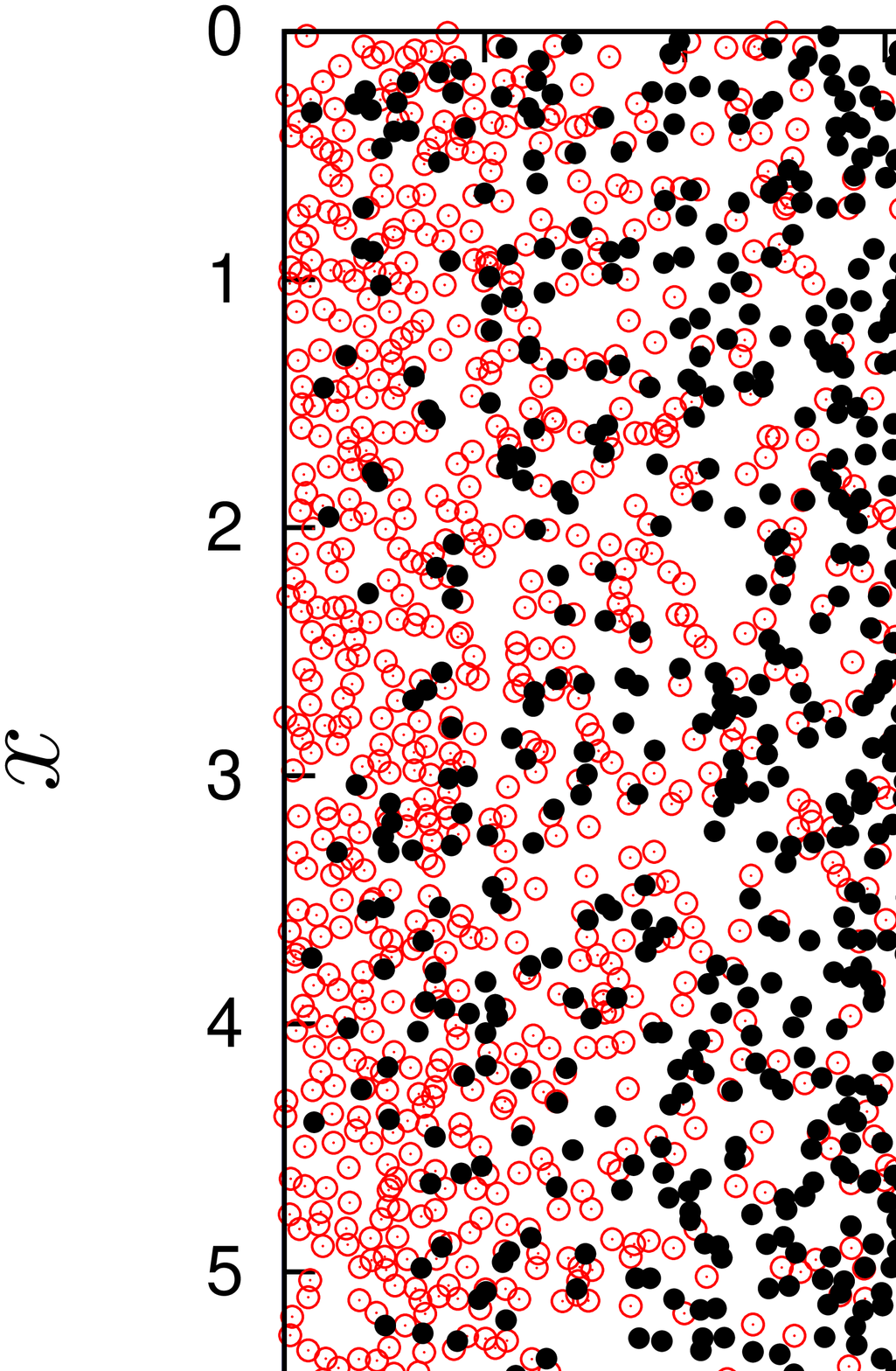}                           
     \end{minipage}\\[5mm]
    \hspace{-3.0cm}
     \begin{minipage}{\textwidth}
      \centering
       \includegraphics[width=0.285\textwidth,height=0.35\textwidth]{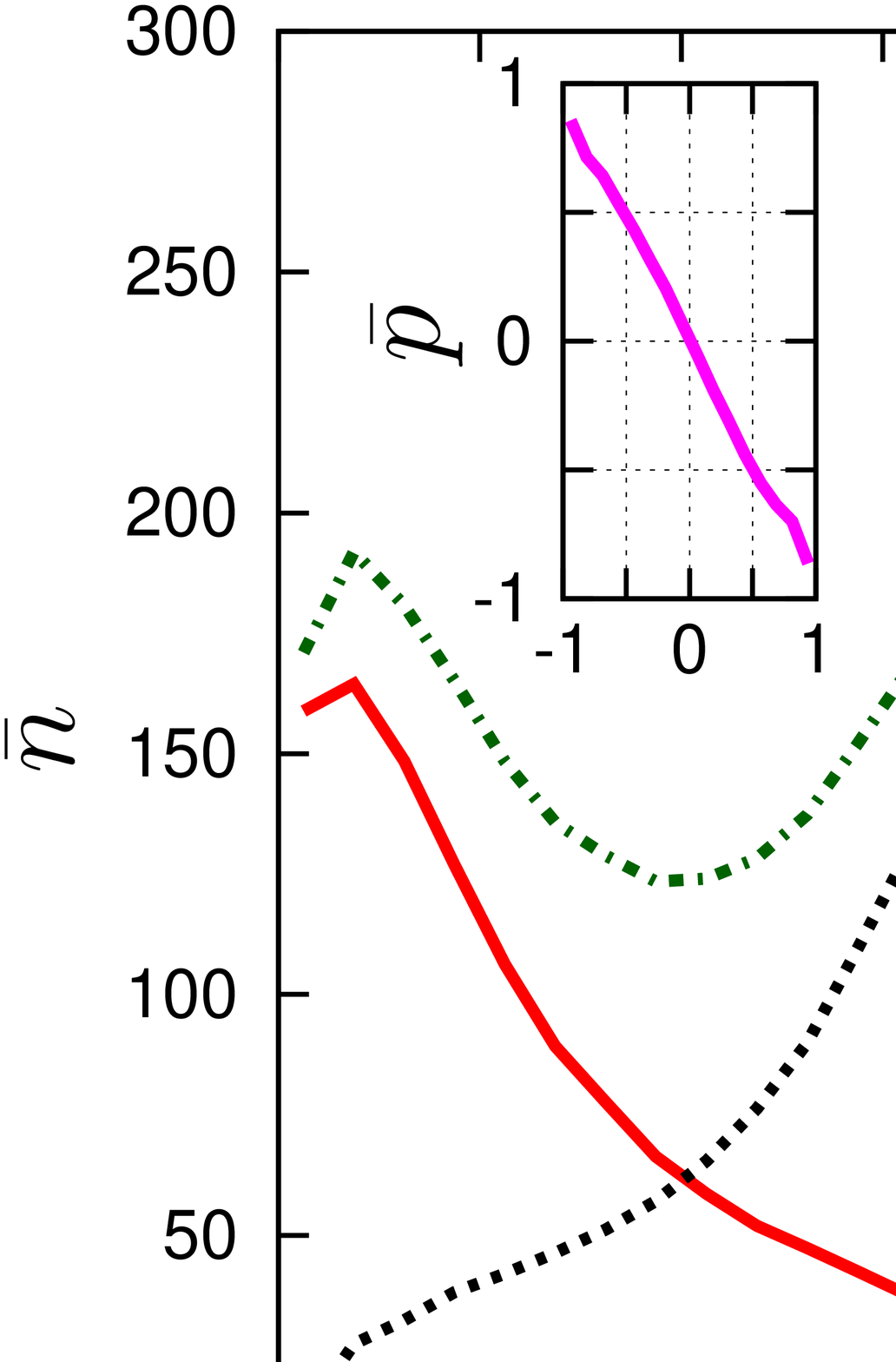}                
     \end{minipage}\\[5mm]
     \hspace{-3.0cm}
     \begin{minipage}{\textwidth}
      \centering
       \includegraphics[width=0.285\textwidth,height=0.35\textwidth]{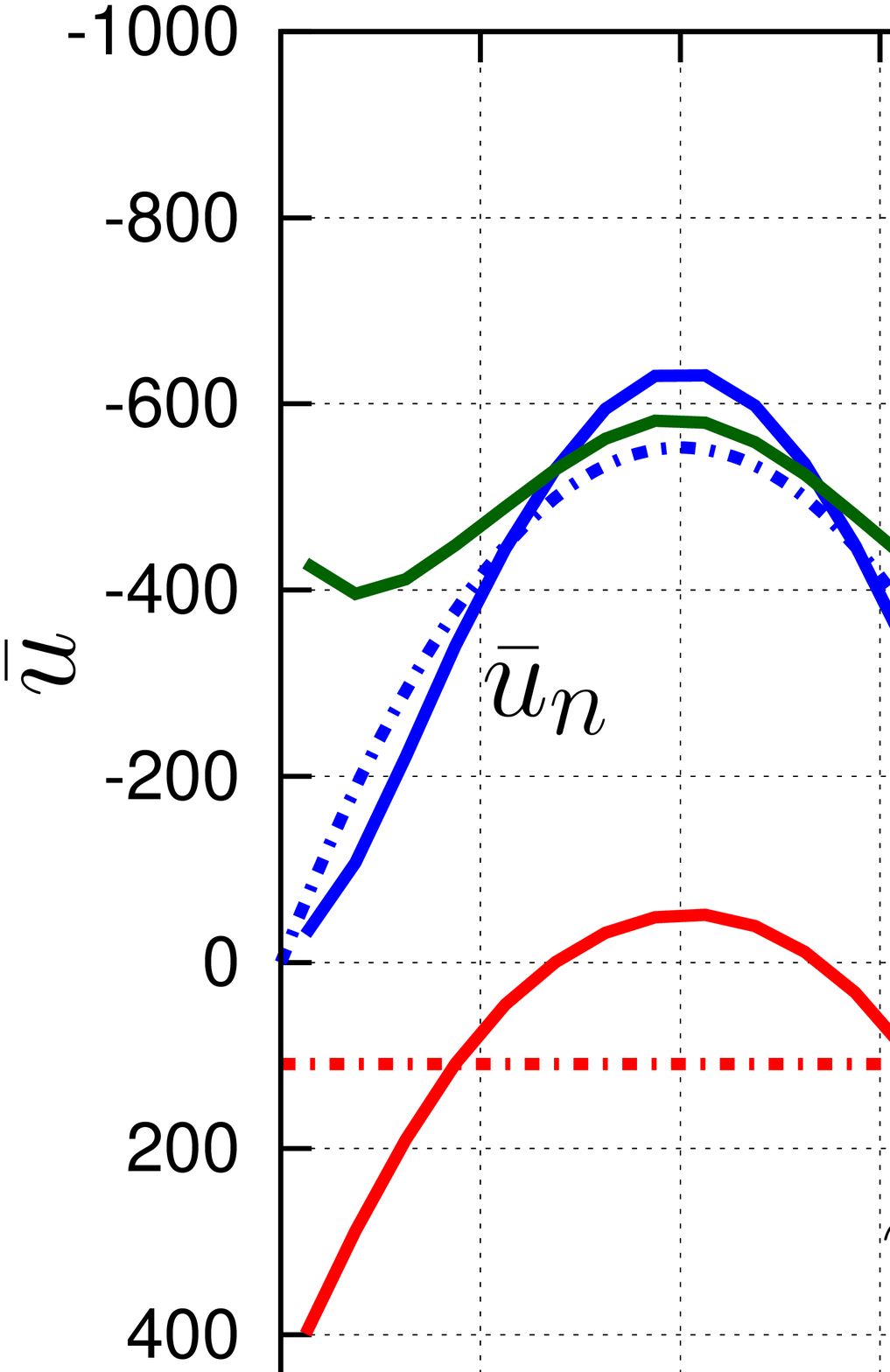}                  
     \end{minipage}
\vspace{1.0cm}
\caption{(Color online). (top): vortex distribution at $t=T_{f}$, red empty (black filled)
 circels indicate positive (negative) vortices; 
(middle): coarse--grained profiles of positive vortex density $\overline{n}^+$ 
(solid red line), negative vortex density $\overline{n}^-$ 
(dashed black line) and total vortex density $\overline{n}$ (dot--dashed green line)
 at $t=T_{f}$. In the inset, the corresponding coarse--grained profile of the 
polarization magnitude $\overline{p}(y)$ is reported (solid magenta line); 
(bottom) coarse--grained profiles of superfluid velocity
$\overline{u}_s$ (solid red line), normal fluid velocity $\overline{u}_n$ 
(solid blue line) and counterflow velocity $\overline{u}_{ns}=\overline{u}_{n}-\overline{u}_{s}$ (solid green line) at $t=T_{f}$. 
Red and blue dot--dashed lines indicate the initial laminar profiles of
the superfluid and the normal fluid, respectively. \label{fig:t=Teq}}
\end{figure}

\begin{figure}[htbp]
\vspace{-1.0cm}
     \hspace{-3.0cm}
     \begin{minipage}{\textwidth}
      \centering
       \includegraphics[width=0.285\textwidth,height=0.35\textwidth]{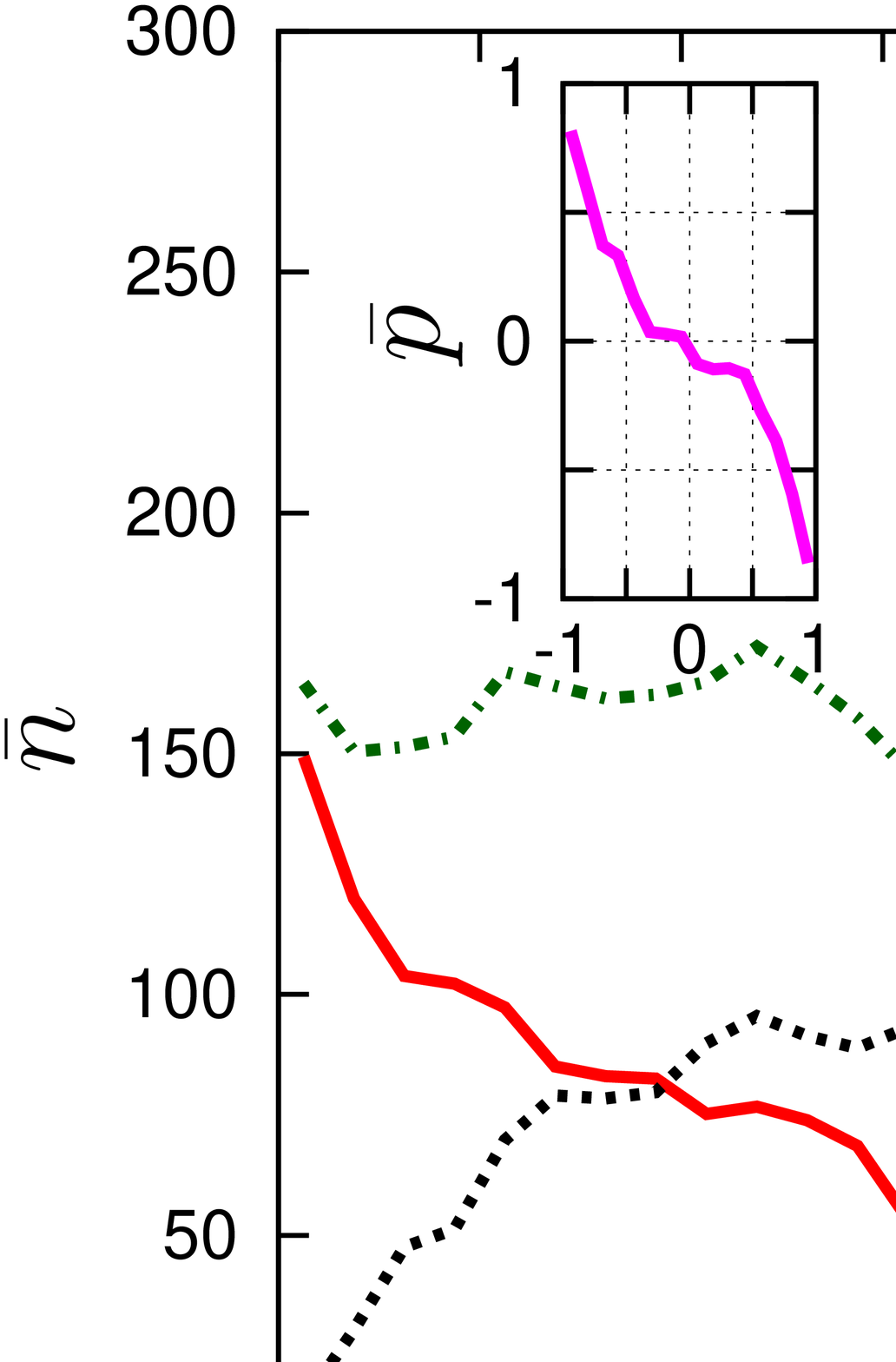}                           
     \end{minipage}\\[5mm]
    \hspace{-3.0cm}
     \begin{minipage}{\textwidth}
      \centering
       \includegraphics[width=0.285\textwidth,height=0.35\textwidth]{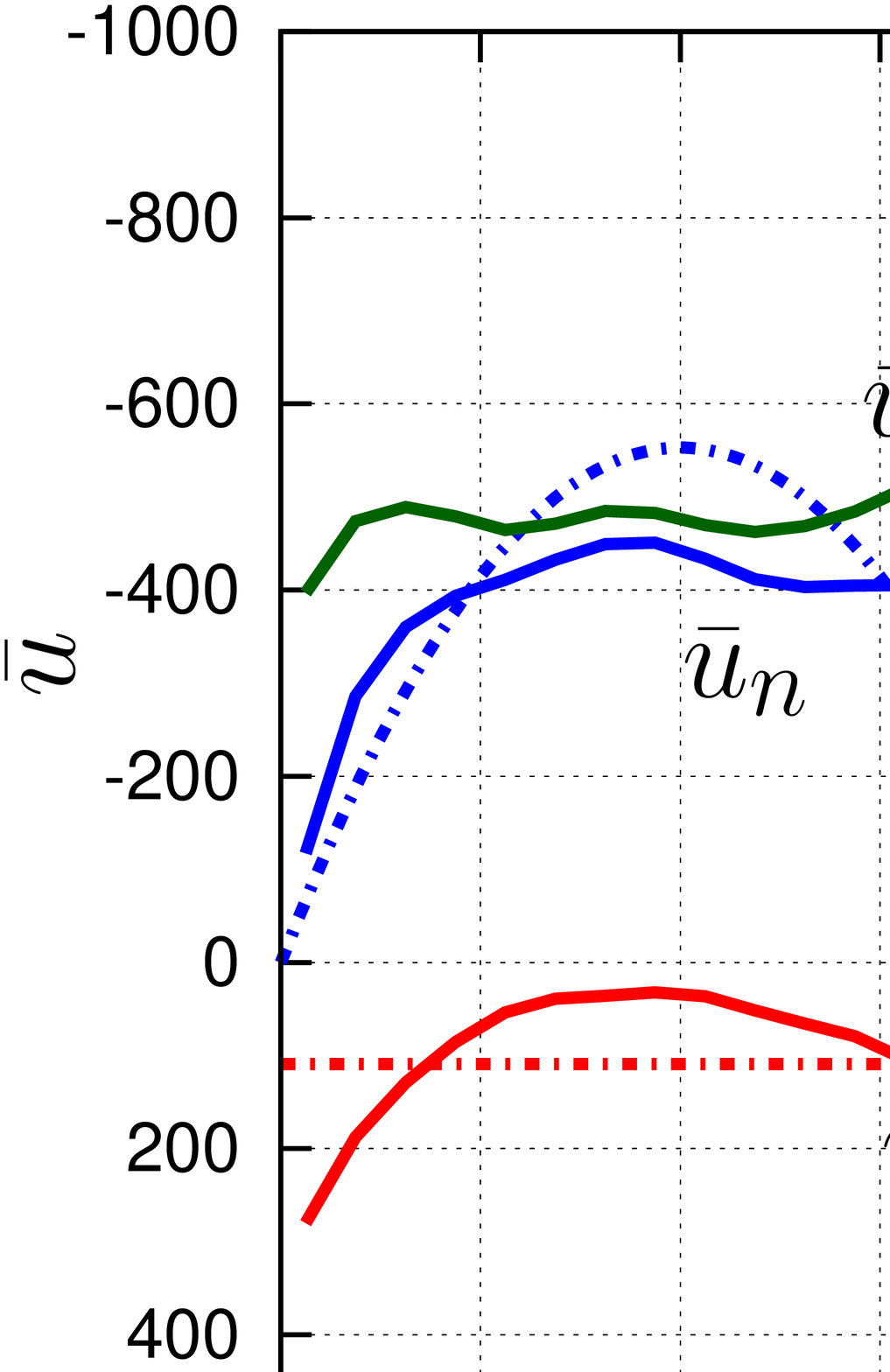}                            
     \end{minipage}
\vspace{1.0cm}
\caption{(Color online). (top): coarse--grained profiles of positive vortex density 
$\overline{n}^+$ (solid red line), negative vortex density $\overline{n}^-$  
(dashed black line) and total vortex density $\overline{n}$ (dot--dashed green line) 
at $t=6.8\times 10^{-3}T_{f}$. In the inset, the corresponding coarse--grained profile 
of the polarization magnitude $\overline{p}(y)$ is reported (solid magenta line); 
(bottom) coarse--grained profiles of superfluid velocity
$\overline{u}_s$ (solid red line), normal fluid velocity $\overline{u}_n$ 
(solid blue line) and counterflow velocity $\overline{u}_{ns}=\overline{u}_{n}-\overline{u}_{s}$ (solid green line) 
at $t=6.8\times 10^{-3}T_{f}$. Red and blue dot--dashed lines 
indicate the initial laminar profiles of
the superfluid and the normal fluid, respectively. \label{fig:t=0.0068Teq}}
\end{figure}

\begin{figure}[htbp]
\begin{center}     
\hspace{-4.0cm}
       \includegraphics[width=0.33\textwidth,height=0.35\textwidth]{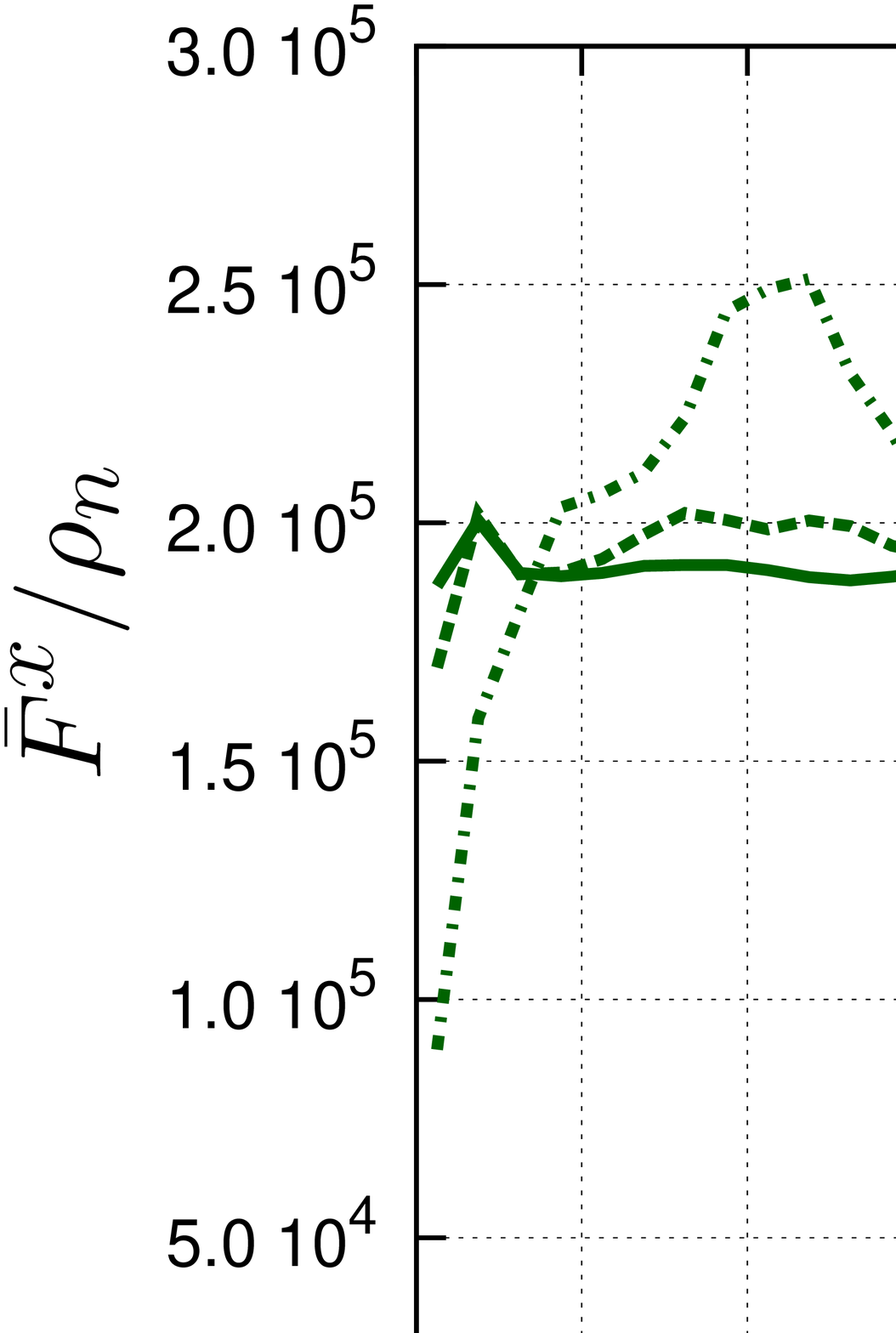}  
\vspace{1.0cm}
\caption{(Color online). Coarse--grained profile of the longitudinal component of the mutual friction force $\overline{F}^x$ 
at different selected times: $t=0$ (dot--dashed green line); 
$t=6.8\times 10^{-3}T_{f}$ (dashed green line); 
$t=T_{f}$ (solid green line).\label{fig:Fns}}
\end{center}
\end{figure}

The evolution to the steady--state can be understood using
the coarse--grained profile of the 
longitudinal component of the mutual friction force $\overline{F}^x$, reported in Fig.~\ref{fig:Fns}. 
The expression of $\overline{F}^x$  at a first order of accuracy according to Eq. \refeq{eq:Fns^pq}, is
\be\label{eq: Fnsx}
\dsp
\overline{F}^x(y)\simeq-\alpha\rho_s\kappa \overline{n}(y)\left ( \overline{u}_n(y)-\overline{u}_s(y)\right )
\ee
At $t=0$, $\overline{F}^x$ is stronger in the central region 
of the channel, flattening the profile of the normal fluid at time 
$t_1\simeq 6.8\times 10^{-3}T_{f}$ very close to the initial configuration, as 
illustrated in Fig.~\ref{fig:t=0.0068Teq} (bottom). 
At times $t\simeq t_1$, the superfluid polarization is 
only partial, see Fig.~\ref{fig:t=0.0068Teq} (top), generating a less pronounced superfluid velocity 
profile $\overline{u}_s(y)$ (Fig.~\ref{fig:t=0.0068Teq} (bottom)). 
The resulting longitudinal component of the mutual friction force at $t\simeq t_1$ 
is therefore more uniform across the channel with respect to $t=0$, as illustrated in Fig.~\ref{fig:Fns}. This allows the normal fluid 
to regain a quasi parabolic profile in the subsequent time interval ($\overline{u}_n$ is approximately parabolic at $t\simeq 0.25 T_f$). 
Finally, at $t=T_f$, the flow reaches a self--consistent dynamical equilibrium determined by 
(a) the vortex--density and velocity profiles reported in Fig.~\ref{fig:t=Teq} (middle) and (bottom) and 
(b) the longitudinal component of the mutual friction force illustrated in Fig.~\ref{fig:Fns}, 
characterized by peak values in the near--wall region.

\section{Discussion\label{sec: Discussion}}
The aim of the present section is to \textit{(a)} describe the idealized
three--dimensional dynamics which we reckon corresponds
to the two--dimensional vortex--points motion
illustrated in Section \ref{subsec: results} and
\textit{(b)} critically discuss to what extent this idealized
three--dimensional motion is capable of grasping the most relevant vortex--tangle
dynamics occurring in helium II T-I counterflows. 
These two issues will be addressed in Sections \ref{subsec: streamwise flow} 
and \ref{subsec: 3D corrispondence}, respectively.
  
\subsection{Streamwise flow of expanding vortex--rings\label{subsec: streamwise flow}}
\begin{figure}[htbp]
\begin{center}     
\hspace{-4.0cm}
       \includegraphics[width=0.3\textwidth]{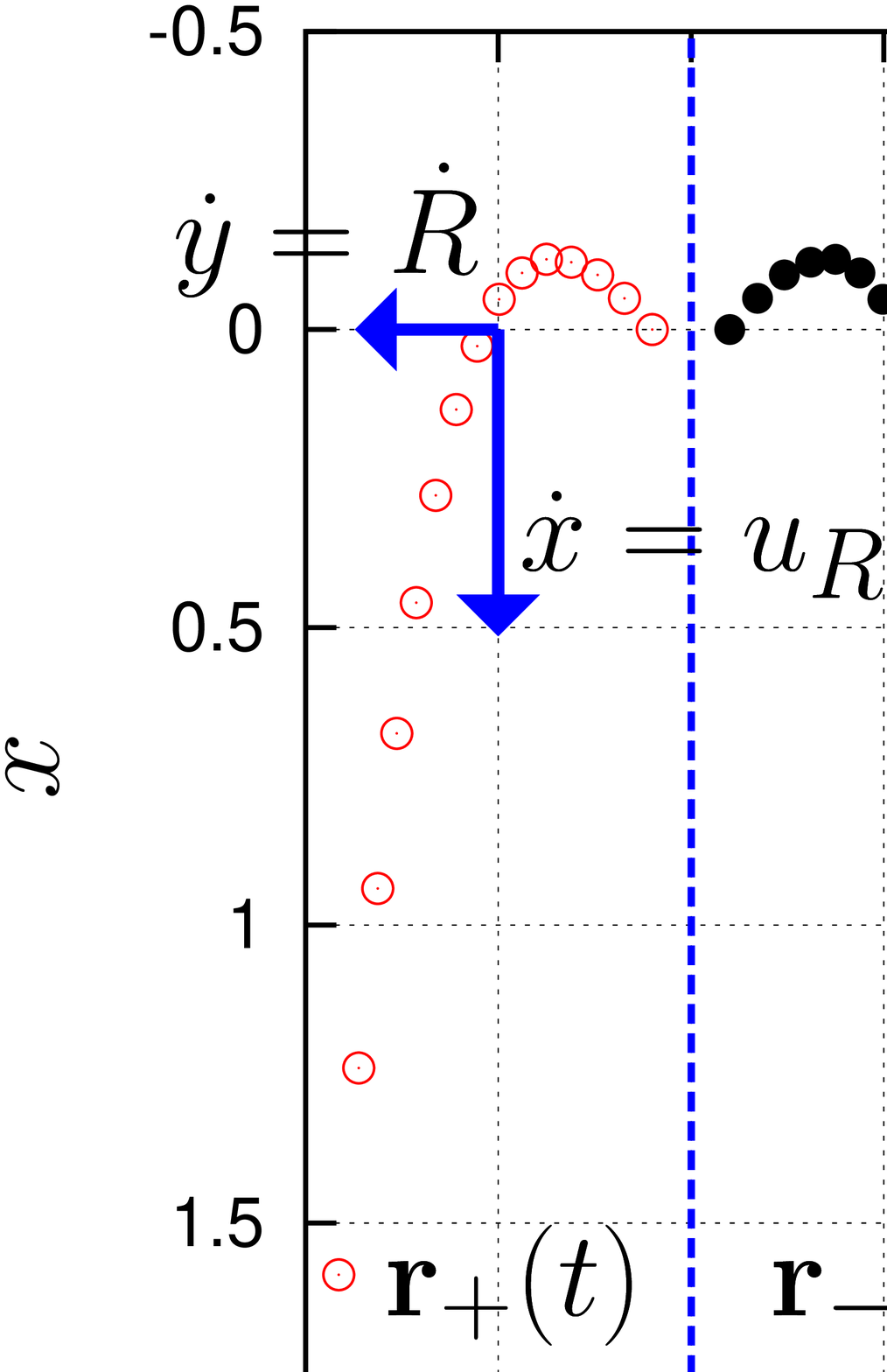}                 
\vspace{1.0cm}
\caption{(Color online). Trajectories $\rr_+(t)$ and $\rr_-(t)$ of an anti--vortex pair, whose initial configuration is symmetric 
with respect to the mid plane of the channel (in dashed blue line). The time interval between consecutives positions is constant, indicating
an increasing streamwise velocity as the vortex--points approach the walls. In an axisymmetric interpretation, 
the three--dimensional analogue of this two--dimensional motion is a streamwise flow of an expanding vortex ring.
\label{fig: anti_vortex}}
\end{center}
\end{figure}

The two--dimensional vortex--points motion described in Section \ref{subsec: results},
can be physically interpreted in three dimensions as an idealized streamwise flow
of expanding vortex--rings lying on planes perpendicular to $\vvn^p$ and drifting
in opposite direction with respect to  $\vvn^p$. This vortex--ring \textit{three--dimensional analogue}
of the vortex--points motion stems from the vortex points equations of motion \refeq{eq:r_j} and can be 
clearly discerned if we consider the motion of an anti--vortex pair whose initial configuration
is symmetrical with respect to the mid plane of the channel and very close to the latter, see Fig. \ref{fig: anti_vortex}.

Let $\rr_\pm (t) = \left ( x_\pm(t), y_\pm(t) \right )$ be the trajectories of the positive and negative vortices 
which consititute the anti--vortex pair, with initial condition $\left ( x^0_\pm, y^0_\pm \right )$.
The axisymmetric hypothesis imposes $\left | y^0_- \right |=\left | y^0_+ \right |=y^0$, 
while the proximity to the channel's mid plane implies $y_0\ll 1$. 
According to the proposed parallel, the dynamics of this anti--vortex pair 
corresponds to the three--dimensional motion of a very small circular vortex ring
centered on the channel's mid plane and initial radius $R^0=y^0$.
To obtain the typical motion of the pair of anti--vortices
(corresponding to the intersections of the vortex ring with 
the two--dimensional channel),
we average the vortex--points equation of motion \refeq{eq:r_j} 
in the streamwise direction and over time deducing
the following equation for $\dsp \overline{\frac{d\rr_\pm}{dt}}$
\begin{eqnarray}\label{eq:r_j-mean}
\displaystyle
\overline{\frac{d\rr_\pm}{dt}} = \overline{\dot{\rr}}_\pm =
\left (
\begin{array}{c}
\overline{\dot{x}}_\pm(y) \\[2mm]
\overline{\dot{y}}_\pm(y)
\end{array}
\right )
=
\left (
\begin{array}{c}
(1-\alpha')\overline{u}_s(y) + \alpha' \overline{u}_n(y) \\[2mm]
\pm \alpha \left ( \overline{u}_n(y) - \overline{u}_s(y) \right ) 
\end{array}
\right ) 
\simeq 
\left (
\begin{array}{c}
\overline{u}_s(y) \\[2mm]
\pm \alpha \overline{u}_{ns}(y)
\end{array}
\right ) 
\end{eqnarray}
where the dot operator indicates the time derivative and $\overline{u}_{ns}=\overline{u}_n-\overline{u}_s$,
to ease notation. In this simple axisymmetric anti--vortex pair model, $\overline{\dot{x}}_\pm=u_R$ and 
$\overline{\dot{y}}_\pm=\dot{R}$, where $u_R$ and $\dot{R}$ are the averaged vortex--ring streamwise drifting velocity
and its expansion rate, respectively. We therefore have the following relations:
\beqn
\dsp
&& u_R(y) = (1-\alpha')\overline{u}_s(y) + \alpha' \overline{u}_n(y) \label{eq: u_R}\\[2mm]
&& \dot{R}(y) =  \pm \alpha \overline{u}_{ns}(y) \label{eq: R_dot}
\eeqn
From equation \ref{eq: R_dot}, given the plot of $\overline{u}_{ns}(y)$ reported in Fig. \ref{fig:t=Teq},
\textit{i.e.} $\overline{u}_{ns}(y)<0\,\,\forall y$, it clearly emerges that the positive (negative) 
vortex moves towards the $y=-1$ ($y=1$). Hence, only the three-dimensional corresponding 
vortex rings whose circulation is oriented in the same direction of $\vvn^p$ expand, while vortex--rings of opposite 
circulation always shrink. The trajectory of an expanding anti--vortex pair is reported in Fig. \ref{fig: anti_vortex}.

\begin{figure}[htbp]
\begin{center}     
\hspace{-4.0cm}
       \includegraphics[width=0.75\textwidth]{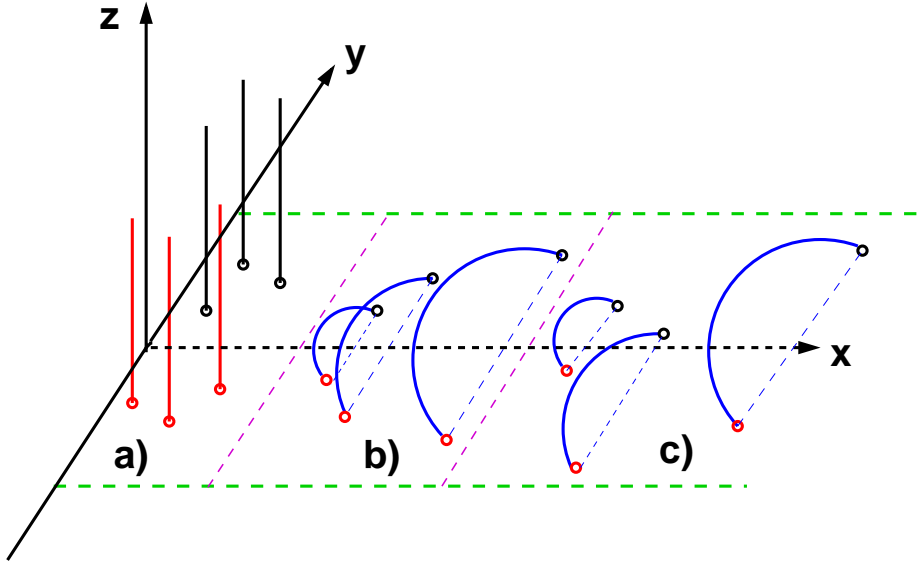}                   
\vspace{1.0cm}
\caption{(Color online). Distinct three--dimensional interpretations of
vortex--points motion: (a) straight vortices model; (b) idealized axisymmetric vortex ring
interpretation described in Section \ref{subsec: streamwise flow}; 
(c) vortex ring analog of the comprehensive two--dimensional motion. \label{fig: 3D_interpr}}
\end{center}
\end{figure}

We would like to stress, however, that our numerical simulations grasp a more general and complex dynamics,
not enforcing an axisymmetric vortex--points motion, but moving each vortex individually.
Therefore, the idealized three--dimensional vortex--ring motion described in the present paragraph is a physical 
interpretation of the average vortex--points motion only. We reckon, nevertheless, that it describes three--dimensionally
the most relevant characteristics of the two--dimensional flow analyzed in the present model. 
In Fig. \ref{fig: 3D_interpr} several three--dimensional physical 
interpretations of the vortex--points motion are illustrated, with Fig. \ref{fig: 3D_interpr} (c) 
describing the vortex ring analog of the comprehensive two--dimensional motion described in 
Section \ref{subsec: results}.

\subsection{Congruity with vortex--tangle dynamics \label{subsec: 3D corrispondence}}
Helium II counterflows are well known to exhibit anisotropic characteristics: the 
vortex lines tend to lie on planes perpendicular to $\vvn^p$. This can be easily deduced, for instance, 
by the plot of the projection of the vortex--line length in the streamwise direction $\langle \Lambda_x\rangle$ in
\cite{baggaley-laurie-2015} and the plots of the anisotropic parameter $I'$ in \cite{yui-tsubota-2015}. 
As a consequence, we reckon that our idelized vortex--rings--flow model 
is able to capture the dynamics of the most relevant fraction of the vortex--tangle. In addition, 
it is worth emphasizing that the vortex--lines aligned in the streamwise direction (which we neglect in our simplified 
three--dimensional interpretation) are only affected very slightly by the mutual friction interaction which
governs the vortex--tangle dynamics. On the other hand, our model is less reliable in the near--wall region
where the vortex--tangle assumes a more isotropic character.

Furthermore, from Eqs. \refeq{eq: u_R} and \refeq{eq: R_dot} and the plots of $u_s$, $u_n$ and $u_{ns}$ 
reported in Fig. \ref{fig:t=Teq} it is possible to deduce that in the proposed three--dimensional physical 
interpretation of our two--dimensional model, the vortex--rings drifting velocity in the streamwise direction
increases as the radius of the vortex--rings grows (\textit{i.e.} as the vortex--rings approach the channel walls).
This vortex dynamics also emerges from past numerical three--dimensional studies 
\cite{baggaley-laurie-2015,yui-tsubota-2015} which describe the vortex lines moving towards the solid boundaries 
with increasing streamwise velocity in opposite direction with respect to the normal fluid flow.

To conclude this section, it is important to underline that the orientation of the expanding vortex--rings
(circulation in the same direction of the normal fluid mean flow) is responsible for the non--uniform profile
of $u_s$ illustrated in Fig. \ref{fig:t=Teq}: the superfluid velocity field induced by such vortex--rings 
slows down the superflow in the central region of the channel while the image vortices increase the superfluid
velocity near the boundaries. This non--uniform superfluid velocity profile is qualitatively recovered in past 
numerical simulations \cite{yui-tsubota-2015}.

Having described what we propose is the three--dimensional physical interpretation of the vortex--points motion 
numerically investigated in our simulations and having discussed its consistency with the vortex--tangle dynamics
observed in past three--dimensional numerical studies, we reckon that our model, although being two--dimensional,
is capable of grasping the most essential and relevant dynamics taking place in helium II T-I channel counterflows.

\section{Conclusions\label{sec: Conclusions}}

In this work we have performed two--dimensional
self--consistent, coupled
numerical simulations
of helium II channel counterflows with corresponding vortex--line density 
typical of counterflow experiments.\cite{ladner-tough-1979,martin-tough-1983} 

The main features of our model are the presence of solid boundaries and the 
dynamical coupling of vortices and normal fluid. These features make our model 
more realistic than previous investigations, although, due to computational 
constraints we had to use a two--dimensional geometry rather than a 
three--dimensional one. 
We reckon, however, that our model, despite its reduced 
dimensionality, is capable of grasping, at least qualitatively, the most relevant
features of the vortex--tangle dynamics occurring in helium II T-I counterflows.
For instance, the proposed physical three--dimensional interpretation 
of the vortex--points motion
(\textit{i.e.} a streamwise flow of expanding vortex rings)
is qualitatively in agreement with the three--dimensional vortex--lines motion computed 
under prescribed normal fluid flow.\cite{baggaley-laurie-2015,yui-tsubota-2015}
In addition, the vortex density profiles computed three--dimensionally with imposed Poiseuille normal
fluid flow \cite{baggaley-laizet-2013,khomenko-etal-2015,baggaley-laurie-2015}
are consistent with the profiles calculated in our two--dimensional simulations. 
Experimentally, these profiles could be estimated by suitable second sound
attentuation measurements, employing high harmonics waves.

In conclusion, the numerical results achieved in our work confirm the already observed velocity matching \cite{barenghi-hulton-samuels-2002} 
and vorticity locking \cite{samuels-1993,barenghi-samuels-bauer-donnelly-1997,barenghi-hulton-samuels-2002,morris-koplik-rouson-2008}
between the two helium II components. Above all, our numerical model predicts the shape of the profile
of the normal fluid which has been just observed experimentally in channels using laser--induced fluorescence
of metastable helium molecules.\cite{marakov-etal-2015} Furthermore, our results are useful for the interpretation of actual
and future experiments, including pure superflow \cite{babuin-stammeier-varga-rotter-skrbek-2012}
and the motion of tracer particles.\cite{chagovets-vansciver-2011,lamantia-duda-rotter-skrbek-2013b}

\begin{acknowledgments}
LG's work is supported by Fonds National de la Recherche, Luxembourg, Grant n.7745104.
MS acknowledges Universit\`{a} di Palermo 
(under Grant Nos. Fondi 60\% 2012 and Progetto CoRI 2012, Azione d).
LG and MS also acknowledge the financial support by the Italian National Group of Mathematical Physics (GNFM-INdAM). 
\end{acknowledgments}

\newpage

%

\end{document}